\documentclass[journal=nalefd,manuscript=letter]{achemso}

\usepackage[version=3]{mhchem} 
\usepackage{xcolor}

\author{Eszter Papp}
\email{eszter.papp@ttk.elte.hu}
\author{Gábor Vattay}
\affiliation{Department of Physics of Complex Systems, Eötvös Loránd University, H-1053 Budapest, Egyetem tér 1-3., Hungary
}
\author{Carlos Romero-Mu\~{n}iz}
\affiliation{Departamento de F\'{i}sica Aplicada I, Universidad de Sevilla, E-41012 Seville, Spain}
\author{Linda A. Zotti}
\affiliation{Departamento de Física Teórica de la Materia Condensada and IFIMAC, Universidad Autónoma de Madrid, E-28049 Madrid,
Spain}
\author{Jerry A. Fereiro}
\affiliation{School of Chemistry, Indian Institute of Science Education and Research Thiruvananthapuram, Thiruvananthapuram- 695551 Kerala, India}
\author{Mordechai Sheves}
\author{David Cahen}
\affiliation{Department of Molecular Chemistry and Materials Science, Weizmann Institute of Science, 76100 Rehovot, Israel}

\title[]{Experimental Data Confirm Carrier-Cascade Model for Solid-State Conductance across Proteins }

\begin{document}

\newpage
\begin{abstract}
The finding that electronic conductance across ultra-thin protein films between metallic electrodes remains nearly constant from room temperature to just a few degrees Kelvin
 has posed a challenge. We show that a model based on a generalized Landauer formula explains the nearly constant conductance and predicts an Arrhenius-like dependence for low temperatures. A critical aspect of the model is that the relevant activation energy for conductance is either the difference between the HOMO and HOMO-1 or the LUMO+1 and LUMO energies instead of the HOMO-LUMO gap of the proteins. Analysis of experimental data confirm the Arrhenius-like law and allows us to extract the activation energies. We then calculate the energy differences with advanced DFT methods for proteins used in the experiments. Our main result is that the experimental and theoretical activation energies for these three different proteins and three differently prepared solid-state junctions match nearly perfectly, implying the mechanism's validity.\\
 
  \includegraphics[scale=1.6]{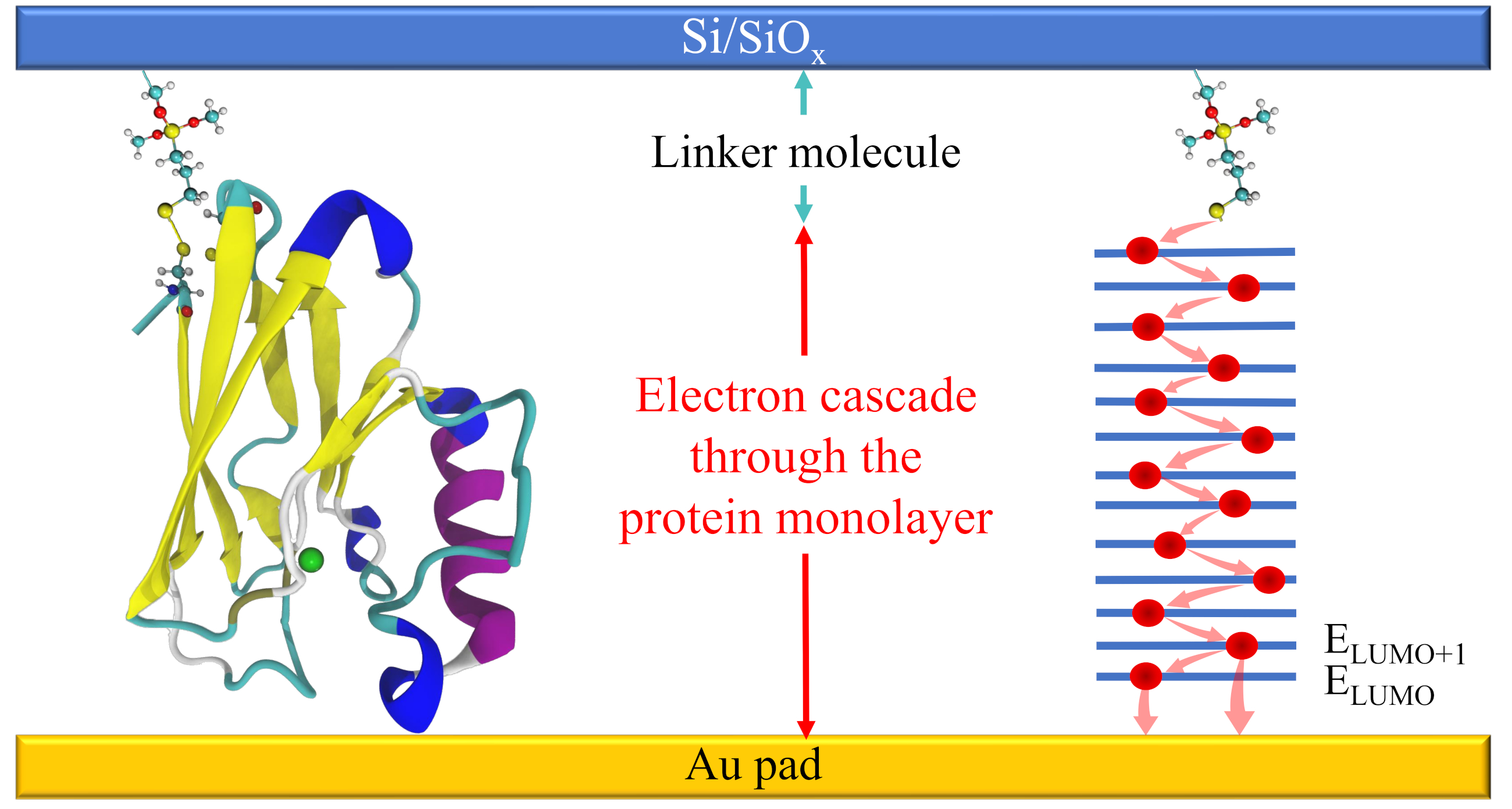}
\end{abstract}

\newpage
\section{Introduction}
Protein-based electronics is an emerging field that has opened the way to developing electronic devices with new functionalities. This is partially due to the extremely efficient charge transfer over relatively large distances these systems exhibit. Many more intriguing observations have been made in experiments incorporating proteins in solid-state devices. Electron {\em transport} measurements via metal-protein-metal junctions show a wide range of anomalous properties compared to electron {\em transfer} in homologous structures\cite{bostick2018protein,amdursky2014electronic,artes2011transistor,zhang2017observation,zhang2019role,yoo2011fabrication,korpany2012conductance,zhang2019electronic}. One of those is that the conductance remains nearly constant when the temperature is changed from $\sim 10$ K to ambient temperatures\cite{sepunaru2011solid, kayser2019solid}.
This seems to indicate that, in order to study these kinds of experiments, a theoretical model capable of describing this low-temperature region is necessary. One of the electron-transport mechanisms which could give rise to temperature independence is coherent tunneling.  Under this assumption, biomolecules chemically bound to metallic contacts can be treated with the Landauer-Büttiker (LB) formalism, which is one of the most common tools used to describe quantum conductance in molecular junctions~\cite{lambert2015basic}. 
It expresses the conductance in terms of the scattering matrix elements between metallic leads. In the simplest case, only a single scattering channel is open in a narrow lead, and a single transmission $T(E_F)$ at the Fermi energy $E_F$ determines the conductance,
$G=G_0T(E_F)$,
where the unit of the quantum conductance is $G_0=2e^2/h$. 
This formula is applicable at zero temperature only. Electron transfer at higher temperatures is usually treated in semiclassical Marcus theory (MT).
Using MT for biomolecules, where the HOMO-LUMO gap is large compared to the thermal energy $k_BT$, Nitzan et al.~\cite{segal2000electron,nitzan2001electron,nitzan2002relationship} derived a formula,
where the conductance is proportional to the electron transfer rate $k_{ET}$ and its temperature dependence is Arrhenius-like. The activation energy is the energy difference between the HOMO and LUMO orbitals.\\ 
Sowa et al.~\cite{sowa2018beyond, doi:10.1063/5.0004514} modeled a molecular junction with a single electronic level coupled to a collection of normalized vibrational modes and derived a generalized master equation. They showed that both LB and MT could be viewed as two limiting cases, and electron-vibrational (electron-phonon) interactions should be treated more carefully in the intermediate regime. 
We further generalized the LB formula~\cite{papp2019landauer} by including all energy levels of the molecule in the junction. We also simplified the mathematical expressions using approximations assuming that the HOMO-LUMO gap is much larger than the thermal energy $k_BT$, a
condition relevant\cite{lever2013electrostatic} in bioelectronic systems operating at ambient temperatures and below. The new LB formula crosses over to that of Nitzan et al., for sufficiently high temperatures. However, when the coupling is strong between the electronic state of the electrode and the nearest localized state of the molecule, new physics arises. The novelty is that electrons that tunnel into the molecule from the metallic contacts at energies higher than
the LUMO energy, even at zero or low temperatures, can go through a downward cascade via transitions between monotonously decreasing energy levels, sustaining a finite conductance. At each cascade stage, electrons can also escape from these levels and reach the weakly coupled other electrode. The probability of these processes is proportional to the Boltzmann factor $e^{-(E-E_{LUMO})/k_BT}$, where $E$ is the energy level
from which the electron escapes the molecule. This process is then dominated by the largest term $e^{-(E_{LUMO+1}-E_{LUMO})/k_BT}$, which then also
dominates the temperature dependence of conductance at low temperatures. In the case of hole transport, the HOMO orbital plays a similar role, and the
conductance is dominated by $e^{-(E_{HOMO}-E_{HOMO-1})/k_BT}$. Next, we present the experimental and computational results validating this mechanism.

\section{Methods}

We introduce three experimental setups, Junctions I-III, where a protein monolayer is sandwiched between two electrodes, as shown in Fig.~\ref{fig:fig1}a and Fig.~\ref{fig:fig2}a, and the current is measured as the function of the applied voltage between the upper and lower
electrodes at different temperatures. Proteins are strongly bonded to the bottom electrode by a covalent bond, while the upper electrodes are
connected weakly.  Holo-Azurin, Apo-Azurin, and a Cytochrome C mutant monolayer were used in the experiments. In Junction I Holo-Azurin and Apo-Azurin from \textit{Pseudomonas aeruginosa} have been attached to a high-doped Si/Silicon oxide substrate with a linker molecule on the bottom and Au pads with lift-off float-on (LOFO) technique on top of the protein monolayer. This experiment has been published in Ref.~\cite{sepunaru2011solid}, and we reanalyze the data. Junction II is the same as Junction I, except that the surface treatment time of the substrate was shorter, resulting in a much stronger coupling between the protein and the electrode. This experiment has not been published before, and the data are analyzed here for the first time. In Junction III CytC (E104C), a monolayer of a mutated Cytochrome of \textit{Equus ferus caballus} has been attached to an Au surface on the bottom, and Au nanowires have been used as top electrodes. Junctions II and III were designed deliberately to increase the coupling between the metallic bottom electrodes and the protein relative to Junction I. This experiment has been published in Ref.~\cite{fereiro2019solid}, and we reanalyzed the data. Details of the experiments can be found in the Supplementary Information. 

\textbf{Theory.} In our previous work~\cite{papp2019landauer} we generalized the zero temperature Landauer-Büttiker conduction formula
for temperatures relevant in these experiments. As mentioned above, for high temperatures, we recovered the formula derived by 
Nitzan et al.~\cite{segal2000electron,nitzan2001electron,nitzan2002relationship} 
\begin{equation}
    G = \frac{e^2}{kT}e^{-\Delta_{HL}/2k_BT}k_{ET},
\end{equation}
stating that the conductance is
proportional to the electron transfer rate $k_{ET}$ between the donor and the acceptor as defined in Ref.~\cite{nitzan2002relationship}, and $\Delta_{HL}=E_{LUMO}-E_{HOMO}$ is the energy difference between the HOMO and LUMO orbitals of the molecule.
The electron transfer rate is also proportional to the factor $e^{-\Delta_{HL}/2k_BT}$ so that the temperature dependence of the conductance becomes
\begin{equation}\label{hight}
    G = C e^{-\Delta_{HL}/k_BT},
\end{equation}
where the constant $C$ depends on the temperature only algebraically. This Arrhenius law results from charge carrier-limited transport where only electrons thermally excited from the HOMO to the LUMO orbital contribute to the current. 

For low temperatures, we found~\cite{papp2019landauer} that the conductance goes to a temperature-independent constant $G_0$ and the convergence to this
is described by a temperature-dependent term $G_T$
\begin{equation}\label{g_temp_simple}
    G = G_0 + G_T.
\end{equation}
The temperature-independent part is
\begin{equation}
    G_0 = \frac{e^2}{h}\left\{ T_h^B(E_F)\frac{\Gamma_{H}^T}{\Gamma_{H}^B}+T_e^B(E_F)\frac{\Gamma_{L}^T}{\Gamma_{L}^B}\right\},
\end{equation}
where $T_{e/h}^B(E_F) = \sum_n (\Gamma_n^T+\Gamma_n^B)\Gamma_n^B/((E_F-\varepsilon_n)^2+(\Gamma_n^T+\Gamma_n^B)^2/4)$ is the tunneling transmission matrix element~\cite{papp2019landauer} between the protein and the Bottom (B) electrode for electrons ($e$)/holes ($h$), $\Gamma_{H/L}^{T/B}$ are
the electronic couplings between the HOMO/LUMO orbitals and the Top/Bottom electrodes. Here we considered that the top electrode
is weakly coupled to the protein, and direct tunneling is not possible or negligible $T_{e/h}^T(E_F)\approx 0$.
The temperature dependence of the conductance in this temperature range is
\begin{equation}\label{cond_temperature}
    G_T =  \frac{e^2}{h}\left\{T_h^B(E_F)\frac{\Gamma_{H}^T}{\Gamma_{H}^B}\left[ \frac{\Gamma_{H-1}^T}{\Gamma_{H}^T}-\frac{\Gamma_{H-1}^B}{\Gamma_{H}^B}\right]e^{-\frac{\Delta_{H,H-1}}{k_BT}}
    + T_e^B(E_F)\frac{\Gamma_{L}^T}{\Gamma_{L}^B}\left[ \frac{\Gamma_{L+1}^T}{\Gamma_{L}^T}-\frac{\Gamma_{L+1}^B}{\Gamma_{L}^B}\right]e^{-\frac{\Delta_{L+1,L}}{k_BT}}\right\},
\end{equation}
where $\Gamma_{H-1/L+1}^{T/B}$ are the electronic couplings between the HOMO-1/LUMO+1 orbitals and the Top/Bottom electrodes, and $\Delta_{H,H-1}=E_{HOMO}-E_{HOMO-1}$ is the energy difference
between the HOMO and HOMO-1 orbitals, while $\Delta_{L+1,L}=E_{LUMO+1}-E_{LUMO}$ is the difference between the LUMO+1 and LUMO orbitals.
Whether electrons or holes dominate the temperature-dependent part of the conductance depends on the coupling strengths, tunneling probabilities, and the energy differences in the exponentials.

From our experimental setup, we can get the voltage and temperature dependence of the measured current. The microscopic coupling strengths and tunneling transmission matrix elements are not accessible. The activation energies $\Delta_{HL}$, $\Delta_{H,H-1}$, $\Delta_{L+1,L}$ and the transmission matrix elements $T_{e/h}^B(E_F)$ depend on the applied voltage. Based on (\ref{hight})
the current in the high-temperature regime is
\begin{equation}\label{current_MT}
    I = I^{high}e^{-\Delta^{high}/ k_BT},
\end{equation}
where $I^{high}$ and $\Delta^{high}$ depend on the voltage. We can then identify $\Delta^{high}$ with the HOMO-LUMO gap $\Delta_{HL}$.
In the low-temperature regime, the current is
\begin{equation}\label{current}
    I = I_0 + I^{low} e^{-\Delta^{low}/k_BT},
\end{equation}
where $I_0$, $I^{low}$ and $\Delta^{low}$ are voltage-dependent, and we can identify $\Delta^{low}$ either with the energy between the HOMO and HOMO-1 orbitals $\Delta_{H,H-1}$ or with the energy between the LUMO+1 and LUMO orbitals $\Delta_{L+1,L}$.

The crossover between the temperature ranges happens approximately where the high- and low-temperature currents are equal.
Since we expect that $\Delta^{low}\ll \Delta^{high}=\Delta_{HL}$, and $I_0$ is negligible at the crossover point, our estimate of the crossover temperature $T_c$
is $k_BT_c\approx \Delta_{HL}/\log (I^{high}/I^{low})$, where $\log(I^{high}/I^{low})\approx 4-7$ has been found\cite{papp2019landauer}. Then using typical
values for the HOMO-LUMO gaps of proteins, the crossover temperature usually lingers in the range of ambient temperatures, and the low-temperature
range then covers the entire temperature range below that.

\begin{figure}[htb]
    \centering
    \includegraphics[width=0.92\textwidth]{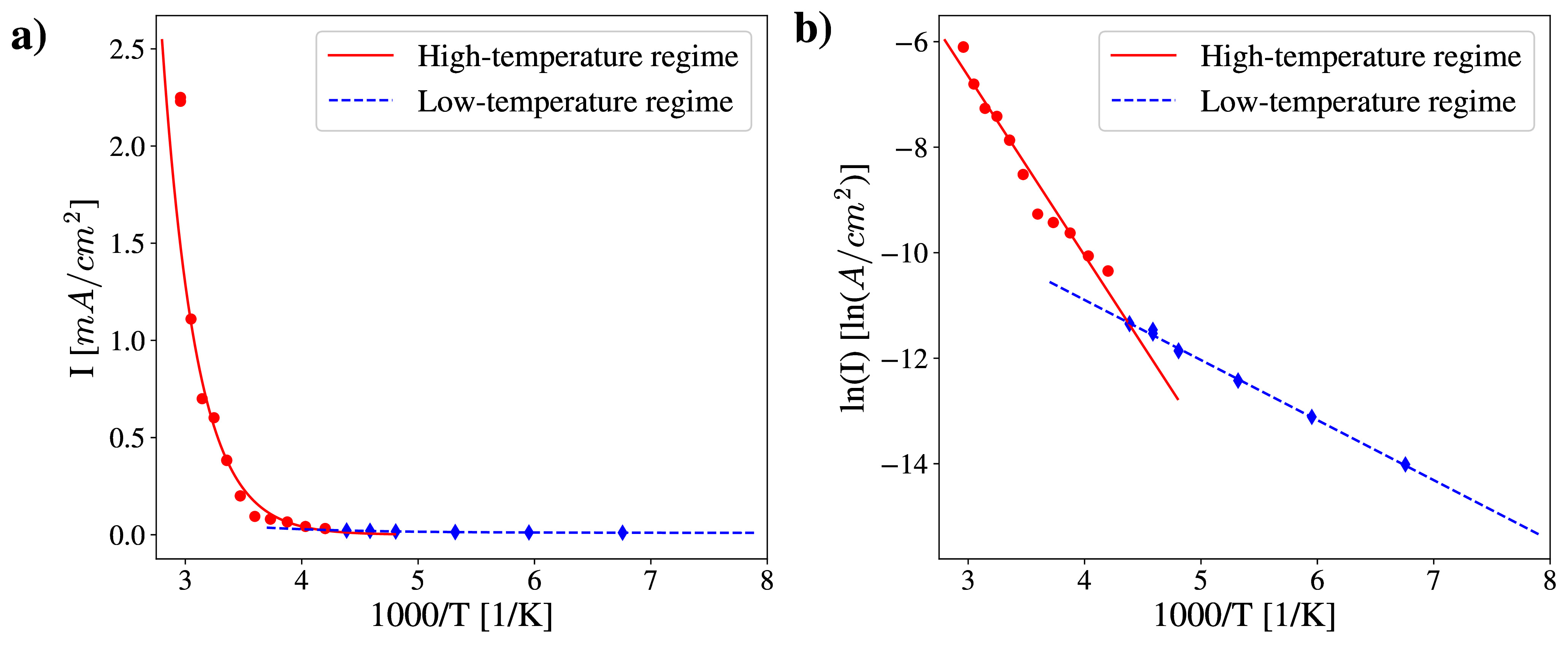}
    \caption{Current density data points of \textit{Junction I} at $-1$ Volts. Data points corresponding to the high- and low-temperature ranges are colored red and blue, respectively. {\bf a)} Raw data points (red and blue dots) as a function of the inverse temperature $1000/T$. {\bf b)} Logarithm of $I-I_0$ for low-temperature data (blue dots) and of $I$ for high-temperature data (red dots) as
    a function of the inverse temperature $1000/T$.  The red and blue lines show the high- and low-temperature theoretical results (\ref{current_MT}) and (\ref{current}), with the best-fit parameters $I$ and $\Delta$. }
    \label{fig:fit_example}
\end{figure}

\textbf{Data analysis.}
We analyzed the temperature dependence of the measured current at fixed voltages. A typical record from Junction I is shown in Fig.~\ref{fig:fit_example}a. In the low-temperature range,
we determined $I_0$ by finding its best value, where $\log(I-I_0)$ as a function of $1/T$ falls on a straight line with the lowest possible regression error. In Fig.~\ref{fig:fit_example}b
we show this in an Arrhenius plot. 

Then both the low and the high-temperature parts fall on straight lines, and the crossover is very sharp between the two. The slopes of the straight lines on the plot are the activation energies of the low- and high-temperature regimes.
We carried out this analysis for all voltages in the experiments. In Fig.~\ref{fig:fig1}e, we show for Junction I, which contained an Apo-Azurin monolayer, the extracted activation energies as a function of the voltage. Then, we repeated the same procedure for Junctions II and III as well. In these two junctions, the high-temperature regime was not present; the sample remained in
the low-temperature regime for the entire temperature range up to ambient temperatures. The parameters of the straight lines on the Arrhenius plot determine the crossover temperature between the low- and high-temperature parts. The absence of the high-temperature regime means that the prefactors in the low-temperature conductance (\ref{cond_temperature}) are large, which is the
consequence of the design of Junctions II and III with strong coupling and high tunneling rate between the bottom electrode and the protein. The extracted low-temperature activation energies
are then shown for Junction II in Fig.~\ref{fig:fig1}d for Holo-Azurin and Apo-Azurin samples and for Junction III in Fig.~\ref{fig:fig2}c for CytC (E104C).

After the determination of the various activation energies from the experimental data, we would like to verify their relationship to $\Delta_{HL}$, $\Delta_{H,H-1}$ and $\Delta_{L+1,L}$
predicted by our theory. To do this, we determined the energies of HOMO, LUMO, HOMO-1, and LUMO+1 orbitals for the proteins involved in the experiments.

\textbf{Electronic structure calculations.}
Technical details about the calculations, including an exhaustive analysis of the charge state and the pH conditions, can be found in our previous works in Refs.~\cite{romeromuniz2018,romeromuniz2019}. Here, we summarize the main points of those calculations.

Density functional theory (DFT) calculations were used to determine the energy and the shape of molecular orbitals of Holo-Azurin, Apo-Azurin and Cytochrome C (E104C) using the efficient OpenMX code. In all cases, we used the geometries obtained from X-ray diffraction experiments available in the Protein Data Bank (PDB). All the structures were processed to add the H atoms in line with the experimental pH conditions, with the aid of the H$++$ tool~\cite{gordon2005}, which can compute pKa values of ionizable side chains of the protein residues. Norm-conserving pseudopotentials were used to replace the Coulomb potential of the core electrons~\cite{morrison1993}, while the Perdew-Burke-Ernzerhof (PBE) functional~\cite{perdew1996} was used to describe the exchange and correlation interaction between valence electrons. The electronic self-consistency was achieved using a Pulay mixing scheme based on the residual minimization method in the direct inversion iterative subspace (RMM-DIIS)~\cite{kresse1996} with a Kerker metric~\cite{kerker1981} and considering an energy cutoff of $10^{-8}$ Ha as a convergence criterion. The OpenMX code is based on highly optimized pseudoatomic orbitals,~\cite{ozaki2003,ozaki2004} which are localized functions employed to represent the single-particle Kohn-Sham wave functions. These pseudoatomic orbitals work with a finite spatial extension given by some cutoff radii that have been set to 8 Bohr for Cu and Fe, 7 Bohr for S, and 5 Bohr for C, N, and H. We have used double-$\zeta$ basis sets for the Cu and Fe metallic centers, while the rest of the atoms were described with single- or double-$\zeta$ basis sets depending on each particular system and atom type. 

In the case of Cytochrome C, we started from the geometry determined for the crystalline structure of horse heart Cytochrome C~\cite{bushnell1990} (1HRC file in PDB). Subsequently, we hydrogenated the system and replaced the glutamate 104 residues with a cysteine amino acid to obtain the E104C mutant used in the experiments.
For the Holo- and Apo-Azurin, we proceeded in a similar way but started from two different input geometries. On the one hand, the geometry obtained for the wild type of Holo-Azurin~\cite{nar1991} and, on the other hand, the apo variant~\cite{nar1992}, that lacks the Cu center (4AZU and 1E65 files in PDB, respectively). In both cases, the minimal structure consists of four protein molecules grouped in a tetramer form. All four monomers possess a similar structure for the Holo-Azurin, while there are noticeable differences between the constituent units for the Apo-Azurin. Namely, the existence of two asymmetrical heterodimers has been reported~\cite{nar1992}. One dimer's molecular structure resembles the one of the Holo-Azurin, while the other is remarkably different. We have analyzed all the possible cases, namely, one monomer of the Holo-Azurin and the two different monomers (labeled as A and B) of the Apo-Azurin. Only Apo-Azurin B is relevant from the point of view of the experiments discussed here. In general, DFT calculations tend to underestimate the HOMO-LUMO gap due to the self-interaction error in the GGA-type calculation. However, our main focus is the low-temperature region where the temperature dependence is related to the $\Delta_{H,H-1}$ and $\Delta_{L+1,L}$ energy gaps, which are not affected by this error systematically. The results of the DFT calculations are summarized in Table~\ref{tab:dft}. The spatial localization of the frontier orbitals, obtained for the gas-phase proteins, are shown in Fig.~\ref{fig:fig1}b-c and Fig.~\ref{fig:fig2}b. Test calculations performed on the Holo-Azurin surrounded by water molecules (not reported here) did not show significant differences as compared to the isolated case.
\begin{table}[htb]
    \centering
    \begin{tabular}{l |c| c| c}  
         & $\Delta_{H,H-1}$ [eV]& $\Delta_{HL}$ [eV] & $\Delta_{L+1,L}$ [eV]  \\
        \hline
         Holo-Azurin & {\color{blue} 0.012} & {\color{blue} 0.369} & {\color{blue} 0.040}\\
         Apo-Azurin (B) & 0.030 & 0.280 & 0.035\\
         Cyt C (E104C) & {\color{red} 0.014} & {\color{red} 0.174 } & {\color{red} 0.085 }\\         
    \end{tabular}
    \caption{The relevant energy gaps from the DFT calculations. Results colored red and blue have been published 
    in Ref.~\cite{fereiro2019solid} and Ref.~\cite{romeromuniz2018}, respectively. Uncolored numbers are published here first.}
    \label{tab:dft}
\end{table}
The voltage dependence of the energy gaps has not been studied here. Therefore we concentrate on the comparison of the zero voltage gaps and 
the corresponding experimental values.

\section{Results and discussion}

\begin{figure}[htbp]
    \centering
    \includegraphics[width=0.9\textwidth]{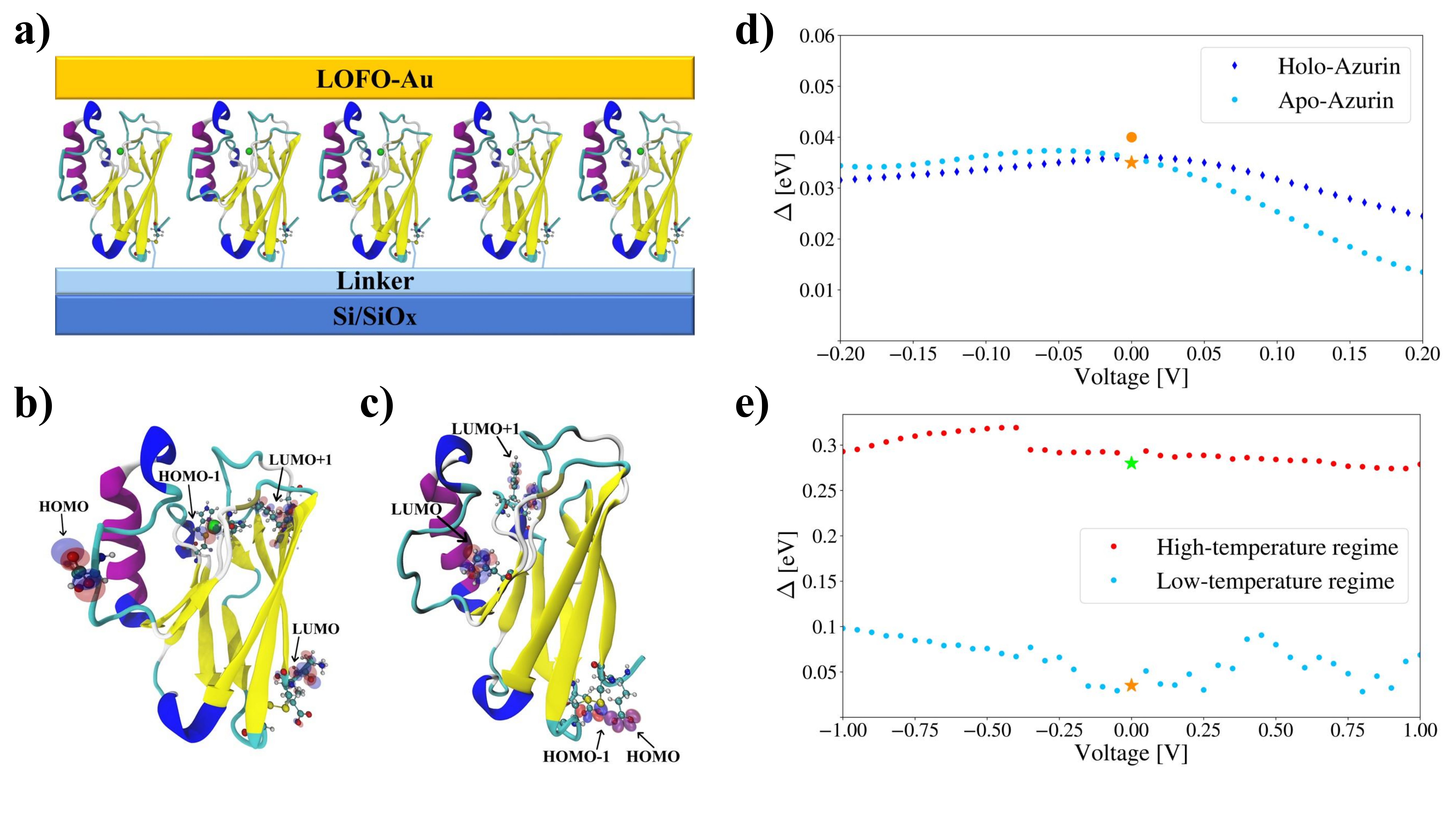}
    \caption{The solid-state Azurin protein junctions. {\bf a)} Schematic illustration of the junction. The ball-stick representation shows the Cu atom (green ball) and the cysteine residues used for binding the proteins to the Si/SiO$_x$ substrate through the linker molecules. In the case of Apo-Azurin, the Cu atom is absent. {\bf b)} Molecular orbitals of Holo-Azurin and {\bf c)} Apo-Azurin. Relevant amino acids are represented by the ball-stick model using CPK coloring.
    {\bf d)} Low temperature $\Delta$ parameters extracted from the experimental data for Junction II for different bias voltages. The orange star and dot are the energy difference between LUMO+1 and LUMO orbitals calculated with the DFT method for Apo-Azurin and Holo-Azurin, respectively. {\bf e)} Low- and high-temperature $\Delta$ parameters extracted from the experimental data for Junction I for different bias voltages. The green and the orange stars are the energy differences calculated with the DFT method for Apo-Azurin between HOMO-LUMO and LUMO+1-LUMO orbitals, respectively. }
    \label{fig:fig1}
\end{figure}

{\bf Junction I.} Apo-Azurin forms the monolayer in this junction, and the bottom electrode receives a long (30 minutes) surface treatment. In this case, the coupling strength between the proteins and the bottom electrode was moderately strong due to the electrostatic barrier caused by the long surface treatment time of the substrate~\cite{garg2018interface}, and as we have shown in Fig.~\ref{fig:fig1}e, the activation energies were extracted for both the high- and low-temperature regimes. Since there is no experimental data point for zero voltage, we determined the zero voltage activation energy by interpolation between the values below and above zero voltage. The zero voltage values extracted from the experiments are shown in Table~\ref{tab:apo_deltas} together with the numerical results of the DFT calculations from Table~\ref{tab:dft} for Apo-Azurin.
\begin{table}[htb]
    \centering
    \begin{tabular}{lcc}
         & \textbf{High temperature} & \textbf{Low temperature} \\
         \hline
          Apo-Azurin experimental $\Delta(0)$ [eV] & $0.291$ & $0.039$\\ 
          Apo-Azurin DFT [eV] & $0.280$ & $0.035$\\
    \end{tabular}
    \caption{The experimental zero voltage $\Delta(0)$ parameters for \textit{Junction I} and the corresponding energy gaps from the DFT calculations for Apo-Azurin.}
    \label{tab:apo_deltas}
\end{table}

In this case, the gap between the LUMO+1 and LUMO dominates the low-temperature behavior. In Fig.~\ref{fig:fig1}e we marked the position of the calculated $\Delta_{HL}$ and $\Delta_{L+1,L}$ values, and one can see that they are in very good agreement with the experimental values subject to experimental variability. As shown in Fig.~\ref{fig:fig1}c, the HOMO and HOMO-1 orbitals overlap at the bottom of the protein, causing good coupling but to the same electrode. As a consequence, $\Gamma_{H}^{B}\approx \Gamma_{H-1}^{B}$ and $\Gamma_{H}^{T}\approx \Gamma_{H-1}^{T}$, and the first term of Eq.~(\ref{cond_temperature}) becomes negligible. This results in an electron-dominated transport with the energy gap $\Delta_{L+1,L}$.

{\bf Junction II.} This junction covers two different experimental setups. In one of them, Holo-Azurin in the other, Apo-Azurin forms the monolayer. The difference between the setups of Junctions I and II is that here
the surface of the bottom electrode receives only a short (3 minutes) treatment, resulting in a better coupling and higher tunneling rate between the protein and the electrode. 
Due to the stronger coupling, the system stays in the low-temperature regime for the entire temperature range of the experiment up to ambient temperatures.
Thus, one can extract only the low-temperature activation energies, which are also shown in Fig.~\ref{fig:fig1}d for the Holo- and Apo-Azurin junctions. 

The zero voltage values extracted from the experiments are shown in Table~\ref{tab:deltas} together with the numerical results of the DFT calculations from Table~\ref{tab:dft} for Holo- and Apo-Azurin.
In this case, the gap between the LUMO+1 and LUMO dominates the low-temperature behavior. In Fig.~\ref{fig:fig1}d we marked the position of the calculated $\Delta_{L+1,L}$ values for Holo- and Apo-Azurin, and one can see that they are in excellent agreement with the experimental values subject to experimental variability. In the case of Holo-Azurin, Fig.~\ref{fig:fig1}b shows that the LUMO and LUMO+1 orbitals are close to the bottom and top electrodes, respectively. Assuming that these unoccupied orbitals are accessible from the Fermi level of the metal contact, their location in space causes a strong coupling between them and the electrodes, making the electron-dominated transport possible, even though the $\Delta _{L+1,L}$ energy gap for electron transport is larger than the $\Delta _{H,H-1}$ for hole transport.

\begin{table}[htbp]
    \centering
    \begin{tabular}{l|c}
         & \textbf{Low temperature} \\
         \hline
          Holo-Azurin experimental $\Delta(0)$ [eV] & $0.036$\\ 
          Apo-Azurin experimental $\Delta(0)$ [eV] & $0.036$\\ 
          Holo-Azurin DFT [eV] & $0.040$\\
          Apo-Azurin DFT [eV] & $0.035$\\
\end{tabular}
    \caption{The experimental zero voltage $\Delta$ parameters for \textit{Junction II} and the corresponding energy gaps from the DFT calculations for Holo- and Apo-Azurin.}
    \label{tab:deltas}
\end{table}

\begin{figure}[htbp]
    \centering
    \includegraphics[width=0.74\textwidth]{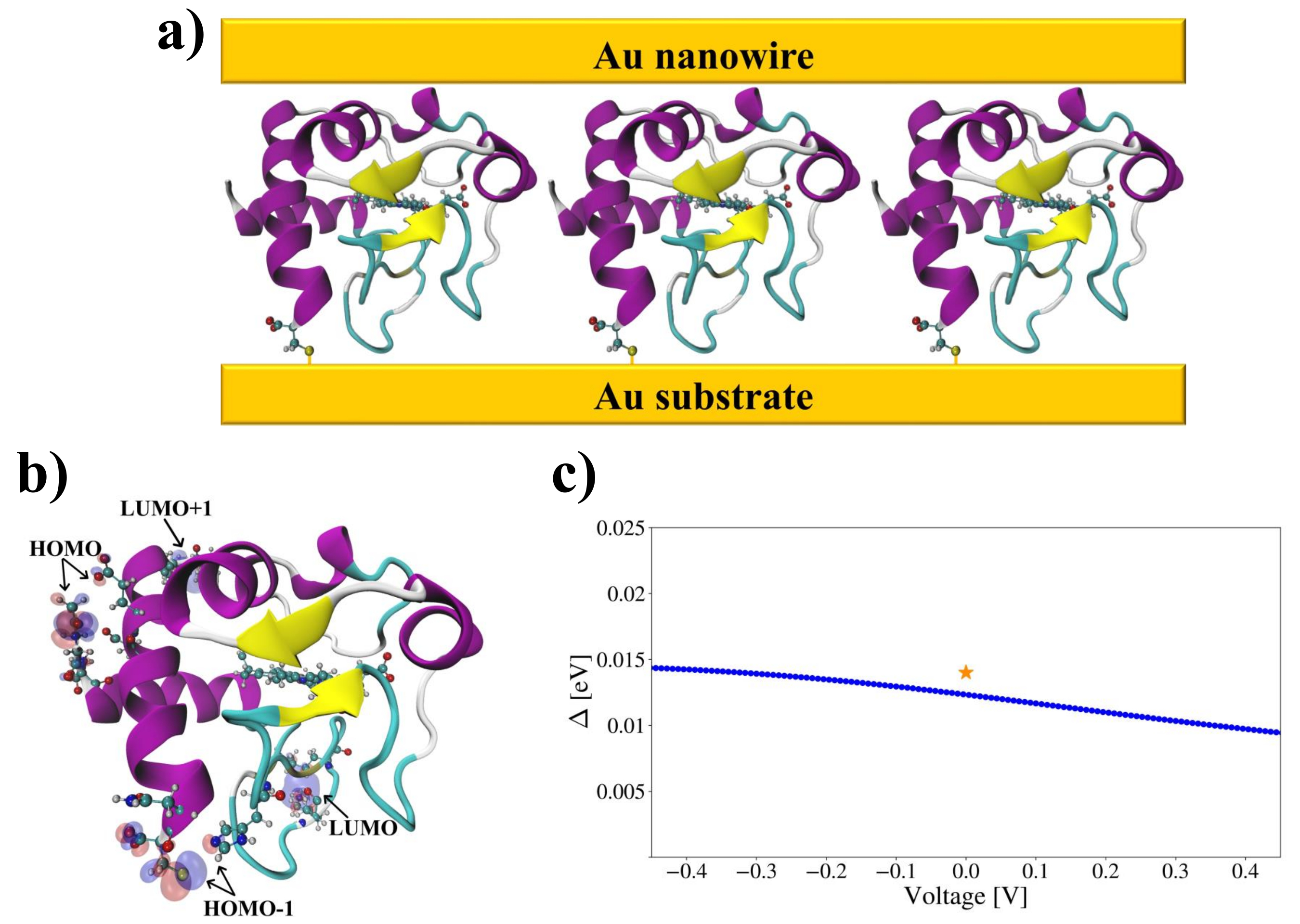}
    \caption{The solid-state Cytochrome C (E104C) protein junction. {\bf a)} Schematic illustration of the junction. The ball-stick representation on the bottom shows the cysteine residues used for binding to the Au substrate. {\bf b)} Molecular orbitals of the protein. The heme group, the acetyl group, and relevant amino acids are represented by the ball-stick model using CPK coloring. The protein connects to the electrode via a bond between its S atom at the bottom and the Au surface. {\bf c)} Low temperature $\Delta$ parameters extracted from the experimental data for Junction III for different bias voltages. The orange star is the energy difference between HOMO and HOMO-1 orbitals calculated with the DFT method. }
    \label{fig:fig2}
\end{figure}

{\bf Junction III. } In this junction, Cytochrome C E104C mutant formed the monolayer. The coupling strength between the proteins and the Au electrodes was strong, so only the low-temperature regime was observed up to ambient temperatures. In Fig.~\ref{fig:fig2}c, the low-temperature activation energies extracted from the experiment are shown. The zero voltage value extracted from the experiments is presented in Table~\ref{tab:C-deltas} together with the numerical result of the DFT calculations from Table~\ref{tab:dft} for Cytochrome C (E104C).
\begin{table}[htb]
    \centering
    \begin{tabular}{l|c}
         & \textbf{Low temperature} \\
         \hline
          Cyt C (E104C) experimental $\Delta(0)$ [eV] & $0.012$\\ 
          Cyt C (E104C) DFT [eV] & $0.014$\\
\end{tabular}
    \caption{The experimental zero voltage $\Delta$ parameters for \textit{Junction III} and the corresponding energy gaps from the DFT calculations for CytC (E104C)}
    \label{tab:C-deltas}
\end{table}

In this case, the gap between the HOMO and HOMO-1 dominates the low-temperature behavior. In Fig.~\ref{fig:fig2}c, we marked the position of the calculated $\Delta_{H,H-1}$ value, which is in good agreement with the experiment. This indicates that hole transport dominates here.
The spatial distribution of the orbitals also supports the assumption of hole-dominated transport. In Fig.~\ref{fig:fig2}b, one can see that the HOMO-1 orbital is on the S atom of the protein’s cysteine residue, which binds it to the bottom Au electrode, coupling this orbital to the electrode strongly. The HOMO orbital is strongly coupled to the top electrode on the opposite side of the protein. The LUMO and LUMO+1 orbitals are also located near the bottom and top electrodes. However, holes dominate the transport process due to the combined effect of the small $\Delta_{H,H-1}$ energy gap and the stronger coupling between the HOMO-1 and the bottom electrode.

\section{Conclusion}
To conclude, we showed that the conductance of proteins strongly bonded to electrodes is, at sufficiently low temperatures, no longer dominated by electrons activated from the HOMO to the LUMO orbitals. Instead, electrons from the electrode can tunnel into spatially close, localized states of the protein, and then
within the protein, they can descend to low-lying states such as the LUMO and LUMO+1. Similarly, holes can ascend to high-lying states such as the HOMO-1 and HOMO. The rate of escape from these final states to the other electrode is the limiting factor of transport. Electrons escaping from LUMO and holes escaping from HOMO maintain a constant zero-temperature contribution to the conductance, which has been experimentally observed first in Ref.~\cite{sepunaru2011solid}. The escape from LUMO+1 and HOMO-1 competes
with this process and modulates the conductance with a temperature-dependent contribution proportional to the Boltzmann factors  $e^{-\Delta_{H,H-1}/k_BT}$ and $e^{-\Delta_{L+1,L}/k_BT}$. In the previous and new experiments presented here, we could identify the low-temperature regime and extract the activation energy from an Arrhenius representation of the data. Using advanced DFT calculations, we showed that the activation energies attained are in excellent agreement with either the gap between the HOMO and HOMO-1 or with the LUMO and LUMO+1 energies depending on the orbitals’ geometry relative to the electrodes. It would be interesting to verify such an agreement by means of more advanced DFT functionals, in order to avoid possible inaccuracies arising from the GGA scheme. Beyond validation of the theoretical background, these results can open
new avenues in understanding other anomalous electron transport properties of proteins\cite{bostick2018protein} and some of the details of electron transport in living systems in general.

\begin{acknowledgement}
The scientific work and results published in this paper were reached with the sponsorship of the Gedeon Richter Talentum Foundation in the framework of the Gedeon Richter Excellence Ph.D. Scholarship of Gedeon Richter. This research was supported by the Ministry of Culture and Innovation and the National Research,
Development and Innovation Office within the Quantum Information National Laboratory of Hungary
(Grant No. 2022-2.1.1-NL-2022-00004). L.A.Z. thanks for the financial support from MCIN/AEI/10.13039/501100011033 (grant PID2021-125604NB-I00) and from the Universidad Autónoma de Madrid/Comunidad de Madrid (grant No. SI3/PJI/2021-00191). C.R.-M. acknowledges funding from the Plan Andaluz de Investigación, Desarrollo e Innovación (PAIDI2020) of Junta de Andalucía. At the Weizmann Inst., this work was supported by a grant from the Israel Science Foundation, the DFG, and, for JF, the Azrieli foundation. M.S. holds the Katzir-Makineni Chair in Chemistry. DC, MS and JF thank D A Dolikh and R V Cherkova, Shemyakin-Ovchinnikov Inst. of Bioorg. Chem., Russ. Acad. Sci, Moscow, for the Cyt C (E104) mutant (see Ref.~\cite{fereiro2019solid}).

\end{acknowledgement}

\newpage

\section{Supporting Information}

\section{Experimental Details}

\subsection{Junction I}

Holo-Azurin (Az) was isolated from Pseudomonas aeruginosa by the method of Ambler and Wynn~\cite{ambler1973amino}.  Apo-Az was prepared by overnight dialysis of $2-3$ mL of Az (1 mg/ml) against 1 L of $0.1$ M KCN (pH 7, adjusted with acetic acid). The dialysis was repeated until no blue color was observed. The cyanide was removed by dialysis against ammonium acetate (0.05 M, pH 8). UV absorption data of Apo-Az showed no absorption peak at 625 nm, indicating the complete removal of the copper ion~\cite{ron2010proteins}.

For the Silicon/Silicon Oxide substrate preparation, batches of p-Si(100) Si wafers were used with specified resistivity of $1.2-1.4$ m$\Omega\,$cm, all with single-side-polished surface and $0.25-0.35$ mm thickness. Before oxide formation, all the wafers were cleaned by successive sonication in acetone/isopropyl alcohol/DI water (3 min in each), followed by 5 min plasma treatment at 100 W with 1:1 Ar:O$_2$. The native oxide was removed by $2\%$ HF treatment for 5 min to leave an active Si-H surface. For controlled growth of a thin oxide layer, the etched Si surface was put in a fresh Piranha solution (H$_2$SO$_4$+H$_2$O$_2$ (3:1)) at 80 $^\circ$C for ${\bf 30\ min.}$ After the treatment, the samples were thoroughly washed with DI water and dried under $N_2$~\cite{garg2018interface, amdursky2015electron}. The oxide thickness was determined using ellipsometry; it was found to be $1.0 \pm 0.1$ nm.

To connect the (Apo-)Az to the Si/SiO$_2$ surface, a self-assembled monolayer of 3-mercap\-to\-propyl trimethoxysilane (3-MPTMS, SH-terminated linker, $97\%$, Sigma-Aldrich) was prepared by sonicating the Si wafer in a $10^{-3}$ M MPTMS in bicyclohexyl solution for 1 hour, followed by 3 min of bath sonication in acetone and 10 s in hot ethanol, yielding a monolayer thickness of $\sim 0.8$ nm. The latter surfaces were then immersed in a vial containing the desired Apo-Az variant for 3 hours. (Apo-)Az was covalently bound to the P$^{++}$ Si/SiO$_x$ substrate via the exposed Cysteine residue (Cys3 or Cys26) that was bound to the 3-MPTMS linker molecule forming the resulting architecture of the sample, as shown in Fig.\,2a. The structure and quality of the produced monolayers were characterized using ellipsometry and AFM (thickness and topography), UV-Vis and polarization modulation infrared reflection absorption spectra (PM-IRRAS).

The top contact was made by placing the depositing 50 nm thick Au pads with the lift-off float-on (LOFO) technique on top of the protein monolayer (see Fig.\,2a). The top contact (Au) was biased, and the back contact was grounded. Thus, both electrical contacts to the protein were electronically conducting and ionically blocking, i.e., no ionic current is measured during what is essentially a DC experiment. The geometric contact area measured by an optical microscope was 0.2 mm in both cases~\cite{ron2010proteins}.

\subsection{Junction II}
The only difference between the first and the second experiment’s preparation is the surface treatment time of the substrate. For Junction I, it was
30 min. and here it was only ${\bf 3\ min.}$ When the treatment time of the surface is long, it becomes less positively charged, which results in an electrostatic barrier~\cite{garg2018interface}, while a short surface treatment time leads to a strong coupling between the protein and the electrode. 

\begin{figure}
    \centering
    \includegraphics[width=0.9\textwidth]{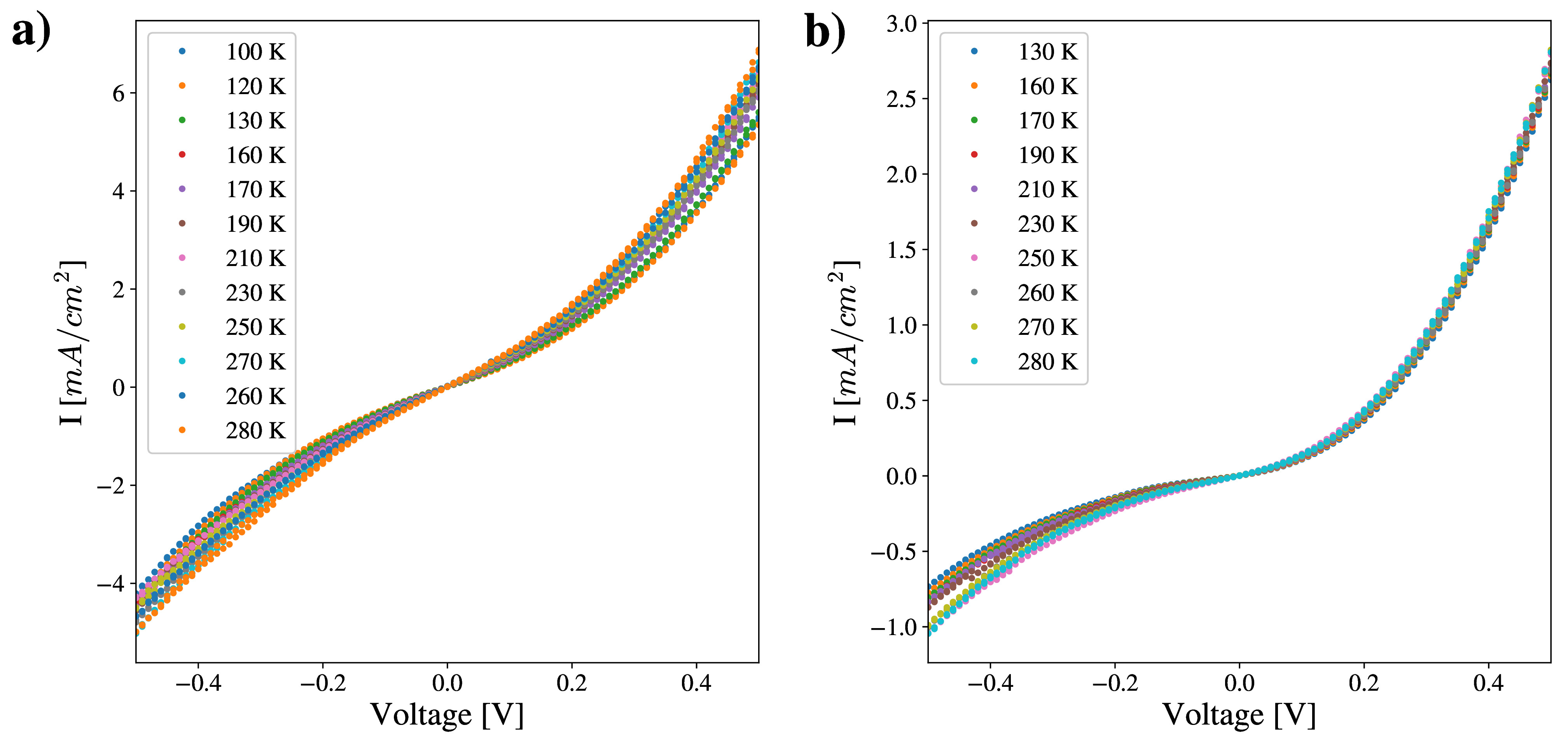}
    \caption{Data from the measurements of Junction II with a monolayer formed by {\bf a)} Holo-Azurin and {\bf b)} Apo-Azurin.}
    \label{fig:measurement_data}
\end{figure}

\subsection{Junction III}

The horse heart genes for the CytC mutants were engineered by R A Chertkova and D L Dolgikh of the Shemyakin-Ovchinnikov Inst. Bio-org. Chem. , Russ, Acad. Sci., Moscow~\cite{fereiro2019solid} using site-specific mutagenesis. The Quik Change Site-Directed Mutagenesis Kit (Stratagene) was used for the experiments. The mutagenesis procedure included the synthesis of the full-length single-stranded DNA plasmid using the thermostable DNA polymerase Pfu, using two complementary primers ($\sim 30-35$ nt) that contained the necessary substitutions. The substitutions were localized in the middle part of the sequence, flanked by $\sim 15$ nt. All of the mutant genes were sequenced to confirm their primary structure. The mutated genes were cloned into the expression system pBP(CYC1) for yeast CytC and modified for horse CytC~\cite{abdullaev2002cytochrome}. The system includes co-expression of the CytC gene and the yeast heme lyase gene. The mutant CytC expression vectors were expressed in Escherichia coli strain JM 109 for $20-22$ hours at 37 $^\circ$C in rich super broth media. The mutant protein was purified from the supernatant obtained after cell disruption by a French press and centrifugation ($100.000\times$g for 20 min at 4 $^\circ$C) using two steps of liquid chromatography: cation exchange on a Mono-S column and adsorption on a hydroxyapatite column~\cite{abdullaev2002cytochrome}. The final purity of the obtained CytC mutant (E104C) was $ >95\%$ according to SDS/PAGE. The expression efficiency of the CytC reached up to 20 mg of the heme-containing CytC per L of culture. The final concentration of the CytC mutant (E104C) was $ \sim1$ mg/mL.

Au electrodes were fabricated on top of a Si wafer using photolithography, yielding a substrate containing 260 devices. The Au electrode surfaces were initially cleaned by bath sonication in acetone/ethanol (3 mins each) and thoroughly rinsed in Milli-Q (18 M$\Omega$) water. The Au surface is activated using ozone (UVO-cleaner Model No: 3422A-220) for 10 mins, followed by treatment with hot ethanol for 20 mins. The activated Au substrates were then rinsed with water and immediately used for incubating with the protein solution.

CytC (E104C) monolayers were prepared by immersing substrates in a 1 mg/mL solution of the protein in 10 mM PBS buffer (pH $\,7.0$) for 3 hours, followed by rinsing it with clean water, followed by drying in a fine nitrogen stream~\cite{fereiro2019solid}. The protein monolayers’ structure and quality were characterized similarly as the (Apo-)Az ones.

A “soft” non-destructive method, the “suspended-wire” technique, was used to form the junctions’ top electric contact~\cite{kayser2018transistor, kayser2019solid, fereiro2020protein}. To this end, an individual Au nanowire (AuNW) with a diameter of $\sim 300$ nm and length of $\sim 5$ $\mu$m is electrostatically trapped by applying an AC bias between the working and reference electrodes, forming a junction between the protein monolayer bound to the lithographically prepared Au (on Si) electrode and the electrostatically trapped single AuNW, as depicted in Fig.\,3a. 

\subsection{Protein thickness measurement}
The thickness of the protein between the two electrodes were measured using
ellipsometry and AFM scratching method.
Ellipsometry measurements were performed with a Woollam
M-2000V multiple wavelength ellipsometer at an angle of incidence of $70^{\circ}$.
AFM scratching method was explained in details elsewhere~\cite{fereiro2021inelastic}. Briefly, An AFM tip was used
in the contact mode (1$\mu$X 1$\mu$) to scratch the surface of the Au surface modified with Protein
monolayers, and then the tapping mode is used (3$\mu$X 3$\mu$) to observe the scratched surface and
determine the thickness of the protein layer using line spectrum. The grooves observed in the
line spectra, carried out at different places in the scratch area of the AFM, indicate the
difference in thickness between the Au surface and the surface modified with proteins. From
these, we can estimate the thickness of the protein’s layer, which is considered as the
thickness between the two electrodes.

\bibliography{cascade}

\end{document}